\documentclass[11pt]{article}
\usepackage{acl2014}
\usepackage{times}
\usepackage{url}
\usepackage{latexsym}
\usepackage{multirow}
\usepackage{graphicx}

\title{Your click decides your fate: Inferring Information Processing and Attrition Behavior from MOOC Video Clickstream Interactions}

\author{Tanmay Sinha$^1$, Patrick Jermann$^2$, Nan Li$^3$, Pierre Dillenbourg$^3$ \\
  $^1$Language Technologies Institute, Carnegie Mellon University, Pittsburgh PA 15213, USA  \\
  $^2$Center for Digital Education, EPFL, CH 1015, Switzerland \\
  $^3$Computer-Human Interaction in Learning and Instruction, EPFL, CH 1015, Switzerland \\
  {\tt  $^1$tanmays@andrew.cmu.edu,  $^2$$^,$$^3$<firstname.lastname>@epfl.ch} \\}

\date{}

\begin{document}
\maketitle
\begin{abstract}
\vspace{-0.1cm}
In this work, we explore video lecture interaction in Massive Open Online Courses (MOOCs), which is central to student learning experience on these educational platforms. As a research contribution, we operationalize video lecture clickstreams of students into cognitively plausible higher level behaviors, and construct a quantitative information processing index, which can aid instructors to better understand MOOC hurdles and reason about unsatisfactory learning outcomes. Our results illustrate how such a metric inspired by cognitive psychology can help answer critical questions regarding students' engagement, their future click interactions and participation trajectories that lead to in-video \& course dropouts. Implications for research and practice are discussed. 
\end{abstract}

\section{Introduction}
\vspace{-0.1cm}
Mushrooming as a scalable lifelong learning paradigm, Massive Open Online Courses (MOOCs) have enjoyed significant limelight in recent years, both in industry and academia \cite{Haggard:13}. The euphoria is about the transformative potential of MOOCs to revolutionize online education \cite{North:14}, by connecting and fostering interaction among millions of learners who otherwise would never have met and providing autonomy to these learners to grapple with the course instruction at their own pace of understanding. However, despite this expediency, there is also considerable skepticism in the learning analytics research community about MOOC productiveness \cite{Nawrot:14}, primarily because of unsatisfactory learning outcomes that plague these educational platforms and induce a funnel of participation \cite{Clow:13}.

With a ``one size fits all" approach that MOOCs follow, scaled up class sizes and lack of face to face interaction coupled with such high student teacher ratios \cite{GuoA:14}, students' motivation to follow the course oscillates \cite{Davis:14}. This is comprehensibly reflected in escalating attrition rates in MOOCs, ever since they have started maturing \cite{Belanger:13,Schmidt:13,Yang:13}. Because it is not feasible for MOOC instructors to manually provide individualized attention that caters to different backgrounds, diverse skill levels, learning goals and preferences of students, there is an increasing need to make directed efforts towards automatically providing better personalized content in e-learning \cite{Sinha:13,Lie:14,SinhaA:14}. The provision of guidance with regard to the organization of the study and regulation of learning is a domain that also needs to be addressed. 

A prerequisite for such an undertaking is that we, as MOOC researchers, understand how diverse ecologies of participation develop as students interact with the course material \cite{Fis:11}, and how learners distribute their attention with multiple forms of computer mediated inputs in MOOCs. Learning in a MOOC requires that students apply self regulation. While substantial research has been done on studying MOOC discussion forums \cite{Ramesh:13,Brinton:13,Anderson:14,SinhaB:14}, grading strategies for assignments \cite{Till:13,Kul:14} and deployment of reputation systems \cite{Coet:14}, inner workings of students' interaction while watching MOOC video lectures have been much less focused upon. Given that roughly 5\% \cite{Huang:14} of students actually participate in MOOC discussion forums, it would be legitimate to ask whether choosing video lectures as units of analysis would be more insightful. After 330,000 registrations in MOOC courses at EPFL in 2013, our experience reflects that out of the 100\% students who register, 75\% show up: 50\% of them primarily watch video lectures and the rest 25\% additionally work out homeworks and assignments. Thus, majority of students have video lecture viewing as their primary MOOC activity.

Video lectures form a primary and an extremely crucial part of MOOC instruction design. They serve as gateways to draw students into the course. Concept discussions, demos and tutorials that are held within these short video lectures, not only guide learners to complete course assignments, but also encourage them to discuss the taught syllabus on MOOC discussion forums. Specific to the context of video lectures, prior work has cut teeth on a)how video production style (slides, code, classroom, khan academy style etc) relates to students' engagement \cite{GuoB:14}, b)what features of the video lecture and instruction delivery, such as slide transitions (change in visual content), instructor changing topic (topic modeling and ngram analysis) or variations in instructor's acoustic stream (volume, pitch, speaking rate), lead to peaks in viewership activity \cite{Kim:14}. There has been increasing focus on unveiling numerous facets of complexity of raw click-level interactions resulting from student activities within individual MOOC videos \cite{GuoC:14,SinhaD:14}. However, to the best of our knowledge, we present the first study that describes usage of such detailed clickstream information to form cognitive video watching states that summarize student clickstream. Instead of using summative features that express student engagement, we leverage recurring click behaviors of students interacting with MOOC video lectures, to construct their video watching profile. 

Based on these richly logged interactions of students, we develop computational methods that answer critical questions such as a)how long will students grapple with the course material and what will their engagement trajectory look like, b)what future click interactions will characterize their behavior, c)whether students are ultimately going to survive through the end of the video and course. As an effort to improve the second generation of MOOC offerings, we perform a hierarchical three level clickstream analysis, rooted in foundations of cognitive psychology. Incidentally, we explore at a micro level whether, and how, cognitive mind states govern the formation and occurrence of micro level click patterns. Towards this end, we also develop a quantitative information processing index and monitor its variations among different student partitions that we define for the MOOC. Such an operationalization can help course instructors to reason how students' navigational style reflects cognitive resource allocation for meaning processing and retention of concepts taught in the MOOC. Our metric aids MOOC designers in identifying which part of the videos might require editing. The goal is to develop an explanatory techno-cognitive model which shows that a set of metrics derived from low-level behaviors are meaningful, and can in turn be used to make effective predictions on high-level behaviors intuitively.

In the remainder of this paper, we describe our study context in the next section. In section 3, we motivate our three level hierarchical MOOC video clickstream analysis (operations, actions, information processing activities), describing relevant related work along the way, along with the technical approach followed. In section 4, we validate our developed methodology by setting up certain machine learning experiments, specifically engagement prediction, next click state prediction, in-video and complete course dropout prediction. Implications for future work and conclusion is presented in section 5.

\section{Study Context}
\vspace{-0.1cm}
The data for our current study comes from an introductory programming MOOC ``Functional Programming in Scala" that was offered on the Coursera MOOC platform in 2012. This MOOC comprises 48 video lectures (10 Gb of JSON data), which has been parsed and preprocessed into a convenient format for experimentation. In these interaction logs, every click of students on the MOOC video player is registered (play, pause, seek forward, seek backward, scroll forward, scroll backward, ratechange). We have information about the rate at which the video lecture is played, total time spent on playing the video and time spent on/in-between various click events such as play, pause, seek etc. In total, 65969 students registered for the course, and 36536 of them had 762137 logged video interaction sessions containing the aforementioned types of click events. If a video is played till the end, then an automatic video-end pause is generated. Otherwise, the Coursera platform unfortunately does not log whether or not a student has left the video in the middle, leaving the true video engagement time unknown. To avoid biased data, we only include video sessions containing video-end pauses. This has yielded a dataset of 222021 video sessions from 21952 students for our analysis in this paper.

\section{Operationalizing the Clickstream}
\subsection{Level 1 (Operations)} From our raw clickstream data, we construct a detailed encoding of students' clicks in the following 8 categories: Play (Pl), Pause (Pa), SeekFw (Sf), SeekBw (Sb), ScrollFw (SSf), ScrollBw (SSb), RatechangeFast (Rf), RatechangeSlow (Rs). When two seeks happen within a small time range ($<$ 1 sec), we group these seek events into a scroll. Additionally, to encode `Rf' and `Rs', we look for the playrate of the click event that occurs just before the `Ratechange' click and compare it with students' currently changed playrate, to determine whether he has sped up/slowed down his playing speed. The reason behind encoding clickstreams to such specific categories, accommodating scrolling behavior and clicks representative of increase and decrease in video playing speed, is to experimentally analyze and understand the impact of such a granularity on our experiments, which are designed with an objective to capture the motley of differently motivated behavioral watching style in students.

As a next step, we concatenate these click events for every student, for every video lecture watched. Thus, the output from level 1 is this string of symbols that characterizes the sequence of clickstream events (video watching state sequence). For e.g: PlPaSfSfPaSbPa.., PlSSbPaRsRsPl..

\begin{table*}[t]
\small
\tabcolsep=0.11cm
\begin{tabular}{|p{3.3 cm}|p{2.9 cm}|p{5.2 cm}|p{3.5 cm}|}
\hline \bf Case (Full, No, Partial match) & \bf Clickstream A & \bf Clickstream B & \bf Fuzzy string matching verdict \\ \hline
{\bf 1:}Varying clickstream length & PlPa{\bf PlSfPaSf}SbSbPl  & PlPa{\bf PlSfPaSf}SbSbPlPaSbSbSbRfRsRf (learner has performed lot more clicks) & Weight(P,A)$>$Weight(P,B)\\ \hline
{\bf 2:}Behavioral pattern appears more than once & PlPa{\bf PlSfPaSf}SbSbPl & PlPa{\bf PlSfPaSf}SbSbPl{\bf PlSfPaSf} \newline (pattern is more characteristic as it appears 2 times) & Weight(P,A)$<$Weight(P,B)  \\ \hline
{\bf 3:}No appearance of behavioral pattern & RfSbSbRs & SSfSSfRsRsRsSfSfSfSfRfRfRfRfRf (string length doesn't matter) & Weight(P,A)$\neq$(P,B)\newline (very low weight) \\ \hline 
{\bf 4:}Variation in number of individual clicks & RfSbSbRs{\bf Pl}Sb{\bf Pa}Sb & RfSbSbRs{\bf Pl}Sb{\bf SfPaSf}Sb \newline (more clicks from pattern appear) & Weight(P,A)$<$Weight(P,B)  \\ \hline
{\bf 5:}Variation in scattering of individual clicks & RfSbRs{\bf Pl}Sb{\bf SfPaSf}Sb (less scattering)  & RfSbRs{\bf Pl}SbSSb{\bf Sf}PlSbRs{\bf Pa}SbRf{\bf Sf} (more scattering) & Weight(P,A)$>$Weight(P,B) \\ \hline
{\bf 6:}Reverse order of individual click appearance & RfRsSb{\bf SfPaSf}Sb{\bf Pl} (order reversed) & RfRs{\bf Pl}Sb{\bf SfPaSf}Sb \newline (order maintained) & Weight(P,A)$<$Weight(P,B)\\ \hline
\end{tabular}
\caption{\label{FZM} Fuzzy string similarity weights for the sample behavioral action P(``PlSfPaSf"). Weight(P, A/B) represents the similarity of the pattern P w.r.t clickstream sequence A or B. }
\end{table*}
\vspace{-0.2cm}
\subsection{Level 2 (Behavioral Actions)} Existing literature on web usage mining says that representing clicks using higher level categories, instead of raw clicks, better exposes the browsing pattern of users. This might be because high level categories have better noise tolerance than naive clickstream logs. The results obtained from grouping clickstream sequences at per click resolution are often difficult to interpret, as such a fine resolution leads to a wide variety of sequences, many of which are semantically equivalent. Therefore, to get more insights into student behavior in MOOCs, we group clicks encoded at very fine granularity into meaningful behavioral categories. Doing this also reduces sequence length which is easily interpretable. There is some existing literature \cite{Ban:00,Wang:13}, that just considers click as a binary event (yes/no) and discusses formation of concept based categories based on the area/sub area of the stimulus where the click was made. 

To summarize a students' clickstream, we obtain n-grams with maximum frequency from the clickstream sequence (a contiguous sequence of `n' click actions). Such a simple n-gram representation convincingly captures the most frequently occurring click actions that students make in conjunction with each other (n=4 was empirically determined as a good limit on clickstream subsequence overspecificity). Then, we construct seven semantically meaningful behavioral categories using these discovered n-grams, selecting representative click groups that occur within top `k' most frequent n-grams (k=100). Each behavioral category acts like a latent variable, which is difficult to measure from data directly.
\begin{itemize}
\itemsep-0.3em
\item {\bf Rewatch}: PlPaSbPl, PlSbPaPl, PaSbPlSb, SbSbPaPl, SbPaPlPa, PaPlSbPa  
\item {\bf Skipping}:SfSfSfSf, PaPlSfSf, PlSfSfSf, SfSfSfPa, SfSfPaPl, SfSfSfSSf, SfSfSSfSf, SfPaPlPa, PlPaPlSf  
\item {\bf Fast Watching}: PaPlRfRf, RfPaPlPa, RfRfPaPl, RsPaPlRf, PlPaPlRf (click group of Ratechange fast clicks while playing or pausing video lecture content, indicating speeding up)
\item {\bf Slow Watching}: RsRsPaPl, RsPaPlPa, PaPlRsRs, PlPaPlRs, PaPlRsPa, PlRsPaPl (click group of Ratechange slow clicks while playing or pausing video lecture content, indicating slowing down)
\item {\bf Clear Concept}: PaSbPlSSb, SSbSbPaPl, PaPlSSbSb, PlSSbSbPa (a combination of SeekBw and ScrollBw clicks, indicating high tussle with the video lecture content)
\item {\bf Checkback Reference}: SbSbSbSb, PlSbSbSb, SbSbSbPa, SbSbSbSf, SfSbSbSb, SbPlSbSb, SSbSbSbSb (a wave of SeekBw clicks)
\item {\bf Playrate Transition}: RfRfRsRs, RfRfRfRs, RfRsRsRs, RsRsRsRf, RsRsRfRf, RfRfRfRf (a wave of ratechange clicks)
\end{itemize}

In an attempt to quantify the importance of each of the above behavioral actions in characterizing the clickstream, we adopt a fuzzy string matching approach. Using this approach, we assign a weight to each of the grouped behavioral patterns for a given students' video watching state sequence (based on similarity of click groups present in each behavioral category, with the full clickstream sequence). The fuzzy string method \cite{Van:14} is justified because it caters to the noise that might be present in raw clickstream logs of students, in six different ways, as mentioned in Table 1. After identifying these cases and meticulous experimental evaluation, we apply the following distance metrics and tuning parameters: Cosine similarity metric between the vector of counts of n-gram (n=4) occurrences for Cases 1 and 2, Levenshtein similarity metric for Cases 3 (weight for deletion=0, weight for insertion and substitution=1), 4, 5, 6 (weight for deletion=0.1, weight for insertion, substitution=1).

As a next step, all subcategories of click groups that lie within each behavioral category are aggregated by summing up the individual fuzzy string similarity weights that are obtained with respect to every students' clickstream sequence. Then, we perform a discretization of these summed up weights, for each behavioral category, by equal frequency (High/Low). The concern of adding up two distance metrics that do not lie in the same range, is thus alleviated, because the dichotomization automatically places highly negative values in the ``Low" category and positive values closer to 0 in the ``High" category. The result is a clickstream vector for each video viewing session of the student, where every element of the vector tells us about the weight (importance) of a behavioral category for characterizing the clickstream. Thus, the output from level 2 is such a summarized clickstream vector. For e.g: (Skipping=High, Fast Watching=High, Checkback Reference=Low, Rewatch=Low, ....).

\subsection{Level 3 (Information Processing)} Watching MOOC videos is an interaction between the student and the medium, and therefore the conceptualization of higher-order thinking eventually leading to knowledge acquisition \cite{Chi:00}, is under control of both the a)student (who decides what video segment to watch, when and in what order to watch, how hard an effort be made to try and understand a specific video segment) and, b)medium/video lecture (the content or features of which decides what capacity allocation is required by the student to fully process the information contained). 

Research has consistently found that the level of cognitive engagement is an important aspect of student participation \cite{Carini:06}. This cognitive processing is influenced by the appetitive (approach) and aversive (avoidance) motivational systems of a student, which activate in response to motivationally relevant stimuli in the environment \cite{Cac:99}. For example, in the context of MOOCs, the appetitive system's goal is in-depth exploration and information intake, while the aversive system primarily serves as a motivator for not attending to certain MOOC video segments. Thus, click behaviors representative of appetitive motivational system are rewatch/clear concept/slow watching, while click behaviors representative of aversive motivational system are skipping/fast watching. In this work, we try to construct students' information processing index, based on the ``Limited Capacity Information Processing Approach" \cite{Bas:94,Lang:96,Lang:00}, which asserts that people independently allocate limited amount of cognitive resources to tasks from a shared pool. Figure 1 depicts this idea.
\begin{figure*}[t]
\includegraphics[scale=0.80]{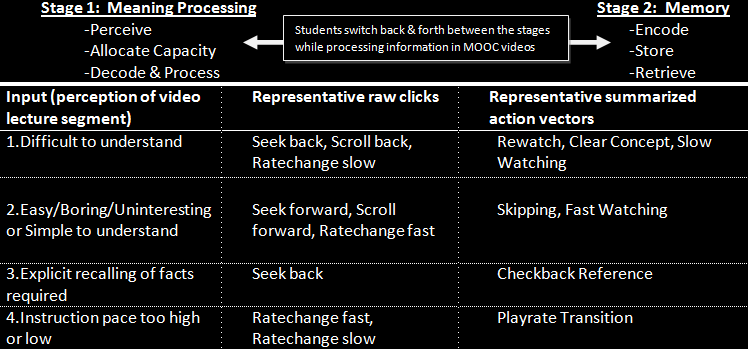}
\caption{Relating students' information processing to click behaviors exhibited in the MOOC, based on video lecture perception}
\label{fig:1}
\end{figure*}

We must acknowledge the fact that video watching in MOOCs requires students to recall facts that they already know (specific chunks of declarative knowledge \cite{Anderson1:14}. This helps them to build a mental representation of the information presented in a MOOC video lecture segment, follow and understand the concept being currently taught. However, it must be noted that depending on the a)expertise level, which decides how available the past knowledge is and how hard is it to retrieve the previously known facts, b)perception of video lecture as difficult or simple to understand, c)motivation to learn or just have a look at the video lecture to seek specific outcomes, cognitive resource allocation would vary among these time sensitive subprocesses in stage 1 and 2 of the pipeline (depicted in Figure 1). This in turn, would be reflected by the underlying non linear navigational patterns that students have, specifically the nature of clicks which they make to adjust the speed of information processing (by pausing, seeking forward/backward, ratechange clicks), as responses to the stimuli. 

Consider an example of students who watch the MOOC lecture, primarily because of reasons such as gaining familiarity with the topic. Such students would purposely not allocate their processing resources to ``memory" part of the information processing pipeline (encode, store, retrieve). Additionally, they will decode and process minimal information that is required to follow the story. On the contrary, students who watch the MOOC lecture, with the aim of scoring well in post-tests (MOOC quizzes and assignments), would allocate high cognitive processing to understand, learn and retain information from the lecture. Thus, such students would process information more fully and thoroughly, despite a possibility of cognitive overload. 

In order to relate our behavioral actions constructed from the raw clickstream with this rich and informative stream of literature, we create a taxonomy of behavioral actions exhibited in the clickstream to construct a quantitative ``Information Processing Index (IPI)". Figure 2 reflects the proposed hierarchy of information processing from high to low using linear weight assignments. We omit the line of reasoning that goes behind defining the precise position of each behavioral action in this hierarchy due to lack of space. However, the details can be found in \cite{SinhaC:14}. Negative weights are necessary to distinguish between ``high" and ``low" weights for each behavioral action. For example, if skipping=high is weighted -3, skipping=low will be weighted +3 on the information processing index. Students' information processing index is defined as follows:\newline{\bf Information Processing Index (IPI)} = \newline $(-1)^j\sum\limits_{i=1}^7$ WeightAssign(Behavioral Action i), \newline {\em j=1,2 depending on whether the behavioral action is weighted low or high}.
\vspace{-0.3cm}
\begin{figure}[h]
\includegraphics[scale=0.75]{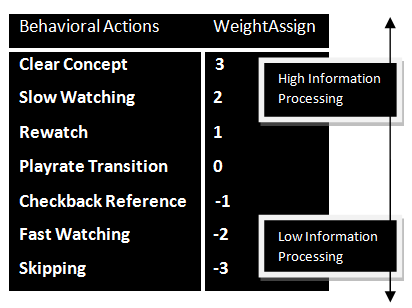}
\caption{Linear weight assignments for behavioral clickstream actions, according to the information processing hierarchy developed}
\label{fig:2}
\end{figure}
\vspace{-0.2cm}

One of the focal utilities of developing such a quantitative index is that meaningful intervention could be provided in real time to students, as they steadily build up their video watching profile while interacting with MOOC video lectures. Viewing throught the lens of the Goldilocks principle \cite{Kidd:12}, our metric can potentially help instructors in understanding and differentiating between students looking away from the MOOC visual sequence, because of too simple or too complex representation. Adaptive presentation of instructional materials is another learning science application where leveraging our metric would be beneficial. 

Specifically, when IPI $>$ 0, it can be inferred that high information processing is being done by students. Therefore MOOC instructors need to check for coherency in pace of instruction delivery and students' understanding. This might also hint towards redesigning specific video lecture segments and simplifying them so that they become easier to follow. On the contrary, when IPI $<$ 0, low information processing is being done by students. Therefore MOOC instructors need to help students better engage with the course, by providing them additional interesting reading/assignment material, or fixing video lecture content such that it captures students' attention. The neutral case of IPI = 0 occurs when students' locally exhibited high and low information processing needs in their evolving clickstream sequence counterbalance each other. So, interventions need to made depending on the video lecture segment, where IPI was $>$0 or $<$0.

\section{Validation Experiments}
We use machine learning to validate the methodology developed in section 3.1 and 3.2 for summarizing students' clickstream, ensuring that the same student does not appear in the train and test folds. The motivation behind setting up these experiments is to automatically measure students' length of interaction with MOOC video lectures, understand how they develop their video watching profile and discern what viewing profile of students leads to in-video and course dropouts. Furthermore, we validate the methodology developed in section 3.3 by statistically analyzing variations of IPI and testing its sensitivity to student attrition using survival models.
\subsection{Machine Learning Experiment Design}
\subsubsection{How much do you engage?}
Students, while watching MOOC video lectures can pause, seek, scroll and change rate of the video. Thus, it is meaningful to quantify students' engagement as the summation of video playing time, seeks \& pauses, multiplied by the playback rate. For example, if a student plays 700 secs out of a 1000 sec video, pauses 2 times for 100 secs each, at an average play rate of 1.5, he effectively engages with the video for (700+200)$*$1.5=1350 secs. Such an interaction measure multiplied by playback rate, is representative of effective video lecture content covered.

{\bf Research Question 1}: Can students' clickstream sequence predict length of students' interaction with the video lecture?

{\bf Settings}: The data for this experiment comes from a randomly chosen video lecture 4-6 (6th lecture in the 4th week of the course, with not too many initial lurkers and not too many dropouts). For experimental purposes, engagement times for students are discretized by equal frequency into 2 categories (High/Low). The dependent variable is student engagement (High: 1742 examples, Low: 1741 examples). L2 regularized Logistic Regression is used as the training algorithm (with 10 fold cross validation annotated by student-id and rare feature extraction threshold being 2). As features, we extract N-grams of length 4 and 5, sequence length and regular expressions from students' clickstream sequences. In the changed setup, we consider summarized behavioral action vectors (output from level 2) as column features.
\subsubsection{Are you bored or challenged?}
Next, we focus our attention on how clickstream sequences evolve. If we know that students' interaction with the video lecture is going to be for a long time (reflected by high engagement), it could have been the case that they were struggling at the current level of instruction (for example, a high combination of pause/seek backward events). Therefore, if such a phenomenon can be detected in real time video lecture interaction, such learners can be presented with reinforcement course material before moving forward. Alternatively, if we know that students' interaction with the video lecture is going to be for a short time (reflected by low engagement), they could be bored or are quite likely to skip course content forward often. Such students can be presented with advanced study material. However, in order to develop such a real time knowledge model and tailor targeted interventions at students, we need to study the trajectory of click sequence formation.

{\bf Research Question 2}: Can we precisely predict what will be the next sequence of clicks that leads students to different engagement states?

{\bf Settings}: The data for this experiment comes from the same video lecture 4-6 (6th lecture in the 4th week of the course). The dependent variable is next click state of students (Pa, Pl, Sf, SSf, Sb, SSb, Rf, Rs). L2 regularized Logistic Regression is used as the training algorithm (with 5 fold cross validation annotated by student-id and rare feature extraction threshold being 5). If we want to predict the click at the i\textsuperscript{th} instant, we extract the following features from 0 till (i-1)\textsuperscript{th} instant: a)Engagement with the video lecture as defined for Research Question 1(High/Low); b)Proportion of click events belonging to Pl/Pa/Sf/SSf/Sb/SSb/Rf/Rs (representative of kind of interaction with the stimulus); c)N-grams of length 4,5 and sequence length from students' clickstream sequences. In the changed setup, we consider summarized behavioral action vectors (output from level 2) as column features.
\subsubsection{Will you drop out of the video?}
As students progress through the video, they slowly build up their video watching profile by interacting with the stimulus in different proportions, which in turn depend on their click action sequences. This motivates our next machine learning experiment, which seeks to derive utility from the first two experiments. Navigating away from the video without completing it fully is an outcome of low student engagement. A student is more likely to watch till the end of a video, if the lecture activates his thinking. Thus, it would be interesting to investigate, whether the nature of students' interaction provides us a hint about in-video dropout behavior. Prior work has made a preliminary study on how in-video dropout is correlated with length of the video, and how in-video dropout varies among first time watchers and rewatchers \cite{GuoC:14}. However, we consider video interaction features at a much finer granularity, representative of how students progress through the video. In doing so, we use detailed clickstream information, including seek, scroll and ratechange behavior, in addition to merely play and pause information.

{\bf Research Question 3}: What video watching profile of students leads to in-video dropouts?

{\bf Settings}: The data for this experiment comes from the same video lecture 4-6 (6th lecture in the 4th week of the course). The dependent variable is the binary variable, in-video dropout (0/1). To address the skewed class distribution, cost sensitive L2 regularized Logistic Regression is used as the training algorithm (with 10 fold cross validation annotated by student-id and rare feature extraction threshold being 2). To extract the interaction footprint of students before they drop out of the video, we extract the following features: a)N-grams of length 4,5 and sequence length from students' clickstream sequences; b)Proportion of click events belonging to Pl/Pa/Sf/SSf/Sb/SSb/Rf/Rs (representative of kind of interaction with the stimulus); c)Engagement with the video lecture as defined for Research Question 1(High/Low); e)Last click action before dropout happened; f)Time spent after the last click action was made (discretized by equal frequency to High/Low). In the changed setup, we consider summarized behavioral action vectors (output from level 2) as column features.
\subsubsection{Will you watch videos and stay till the course end?}
We may expect that when students find the course too tough to follow, uninteresting or boring, they will not engage with future videos. On the contrary, when students seem very interested in understanding the video and exhibit lots of rewatching behavior, we might expect them to stay on till the course end video lectures. Students who do not stay till the last week of the course (exhibit any video lecture viewing), are considered as complete course dropouts. One principal application of detecting these dropouts early could be recommendation of selected future video lectures to watch (for example, where an interesting concept, case study or application is going to be discussed), to positively motivate and pull these students back into the MOOC.

{\bf Research Question 4}: Can we discover patterns in the video watching trajectory of students that can predict when are students most likely not to view future video lectures? 

{\bf Settings}: The data for this experiment comes from all 48 videos of ``Functional Programming in Scala" MOOC (4710 non-dropouts, 9596 dropouts). To address the skewed class distribution, cost sensitive L2 regularized Logistic Regression is used as the training algorithm (with 5 fold cross validation annotated by student-id and rare feature extraction threshold being 5). The dependent variable is the binary variable, complete course dropout (0/1), indicating whether the student ultimately stayed on (watched any video lecture) till the last course week. Engagement (time in seconds) of a student is discretized by equal frequency into High and Low categories, considering all interactions in each video lecture separately (because length of each video differs, so the discretization criteria would also differ for each video). Video play proportion((video played length/video length)*100*average play rate) for a student is discretized by equal width (Very Low: $<$50\%, Low: 50-100\%, High: 100-150\%, Very High: $>$150\%). IPI for a student is discretized by equal frequency (Very Low: $<$-1.00, Low: [-1.00, 1.00], High: [1.00, 3.00], Very High: $>$3.00). The discretization criteria (equal width, frequency and number of bins) was experimentally determined. Development of trajectories for each of these factors is indicated in Figure 3. To extract the interaction footprint of students before they drop out of the course, we extract the following features: a)N-grams of length 4,5 and sequence length from ``Engagement trajectory", ``Video Play Proportion trajectory" and ``IPI trajectory" of students for the videos watched from 0 to (n-1)th instant, b)Engagement, Video Play Proportion and IPI trajectories for the nth instance (attribute for the last video lecture watched before dropping out), c)Proportion of different symbol representations in the trajectories (for example, in a trajectory such as HLLHH, proportion(H)=60\%, proportion(L)=40\%. 
\begin{table*}[t]
\small
\begin{tabular}{|p{2.0 cm}|p{1.9 cm}|p{1.3 cm}|p{2 cm}|p{6.7 cm}|}
\hline \bf Research Question & \bf Condition & \bf Accuracy \newline Kappa & \bf False Negative Rate & \bf Most representative (weighted) features that characterize classes \\ \hline
1. Engagement \newline Prediction & \multirow{2}{*} {A)Raw Clicks} & 0.81 \newline 0.63 & {\bf 0.24} & {\bf High} (skipping=low, playrate transition=low, rewatch=high, slow watching=low, checkback reference=low, clear concept=high)\\  
& B)Summarized \newline Behavioral \newline Action \newline Vectors & 0.75 \newline 0.49 & {\bf 0.15} &  {\bf Low} (skipping=high, playrate transition=high, rewatch=low, slow watching=high, checkback reference=high, clear concept=low) \\ \hline
2. Next Click \newline Prediction & \multirow{2}{*} {A)Raw Clicks} & 0.68 \newline 0.57 & - & {\bf SeekFw} (playratetransition=low, skipping=low, fast watching=high, clearconcept=low) \newline {\bf SeekBw} (checkbackreference=high, rewatch=low, playratetransition=low, propSeekBw, clearconcept=high)\\  
& B)Summarized \newline Behavioral \newline Action \newline Vectors & 0.66 \newline 0.54 & - &  {\bf Ratechangefast} (playratetransition=high, rewatch=low, checkbackreference=low) \newline {\bf Ratechangeslow} (playratetransition=high, clearconcept=high) \\ \hline
3. In-video \newline dropout \newline Prediction & \multirow{2}{*} {A)Raw Clicks} & 0.90 \newline 0.69 & {\bf 0.19} & {\bf Non dropouts} (skipping=low, clearconcept=high, slow watching=high, Checkbackreference=low, rewatch=high, engagementfromStart=low, engagementlastClick=high)\\  
& B)Summarized \newline Behavioral \newline Action \newline Vectors & 0.90 \newline 0.70 & {\bf 0.15} &  {\bf Dropouts} (skipping=high, clearconcept=low, slow watching=low, engagementfromStart=high, rewatch=low, engagementlastClick=low, checkbackreference=high) \\ \hline
4. Complete \newline Course dropout \newline Prediction & Operationalized trajectories & 0.80 \newline 0.57 & {\bf 0.143} & {\bf Non dropouts} (trajectory\_IPI=H H H H, trajectory\_eng=H H H VL H, trajectory\_vpp=H H H L H) \newline {\bf Dropouts} (trajectory\_IPI=H  H VL VL VL, trajectory\_eng=H L H L L, trajectory\_vpp=H H H H VL)\\ \hline
\end{tabular}
\caption{\label{FZM} Performance metrics for machine learning experiments. Random baseline performance is 0.5}
\end{table*}
\vspace{-0.2cm}
\begin{figure}[h]
\includegraphics[scale=0.52]{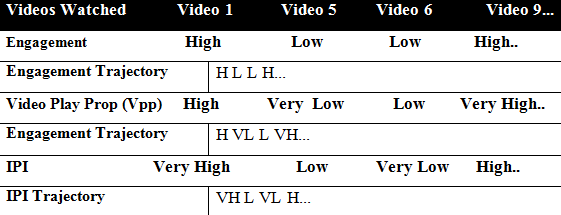}
\caption{Example depicting how different operationalized trajectories of students are formed}
\label{fig:3}
\end{figure}
\vspace{-0.3cm}
\subsection{Results} 
Results of the four machine learning experiments, along with the most representative (weighted) features that characterize classes, are reported in table 2. There are two important positives here: a)The summarized behavioral action vectors from level 2 are able to achieve nearly similar values of accuracy and kappa when compared to the raw level clicks. This means that we can reason different meaningful video viewing behaviors of students without getting our hands dirty in examining noisy and continually occurring raw clicks,  b)Our metric of interest, i.e the false negative rate\footnote{False negative rate of 0.x means that we correctly identify (100-(100*0.x))\% of dropouts} is lower for Case 1.B and Case 3.B, as compared to Case 1.A and Case 3.A, which shows the effectiveness of the clickstream summarization approach (level 2) in pre-deciphering the fate of students to some extent.

Additionally, we leverage a statistical analysis technique referred as survival analysis \cite{Miller:11}, to quantify the extent to which our summarized behavioral clickstream action vectors and IPI are sensitive to students' dropout. In this modeling scheme, dropout variable is 1 on the students' last week of active participation (in terms of video lecture viewing), and is 0 for all other weeks. Our investigation results indicate that a)Students' dropout in the MOOC is 37\% less likely, if they have one standard deviation greater IPI than average (Hazard ratio: 0.6367, p$<$0.001). Such students grapple more with the course material to achieve their desired learning outcomes (as reflected by their video lecture participation), b)If students' rewatching behavior changes from low to high, they are 33\% less likely to dropout (Hazard ratio: 0.6734, p$<$0.001), c)As students start watching more proportion of the video lecture, they are 37\% less likely to dropout of the MOOC (Hazard ratio: 0.6334, p$<$0.001). This is indicative of their continued interest in the video lecture.
\begin{figure*}[t]
\includegraphics[scale=0.60]{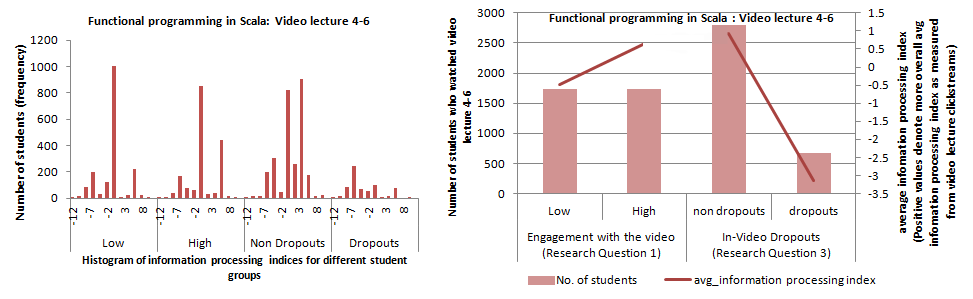}
\caption{Variation of Average Information Processing Indices(IPI) for Video 4-6}
\label{fig:4}
\end{figure*}
\begin{figure*}[t]
\includegraphics[scale=0.60]{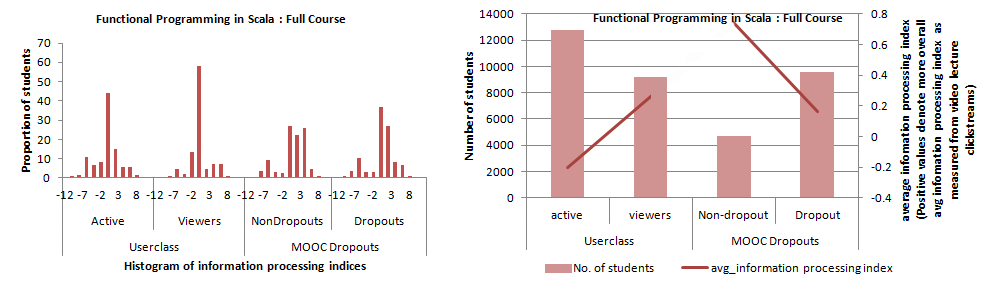}
\caption{Variation of Average Information Processing Indices(IPI) for the full course}
\label{fig:5}
\end{figure*}
\vspace{-0.6cm}

Next, to discern how IPI fluctuates among different student partitions and validate whether our operationalization produces meaningful results, we plot figures 4, 5 and perform statistical tests, specifically z test (testing significance of difference between means for large sample sizes, when population standard deviation is known). Population refers to all students in the MOOC being currently studied. The right half of figure 4 depicts the variation of average IPI, among high versus low engagers and in-video dropouts versus non dropouts, in the same video lecture 4-6 from the course, that we have been performing our experiments on. Similar findings were also confirmed with other randomly chosen course videos. The left half of figure 4 shows the frequency distribution of average IPI. This figure concurs with our intuitions. The average IPI is significantly higher for students with ``High" engagement ($|$z$|$=8.296, p$<$0.01) and ``Non In-video dropouts" ($|$z$|$=22.54, p$<$0.01). This is also reflected in the histogram, which clearly shows that many non in-video dropouts have positive IPI that pushes up the average. Because the effect is smaller in low engagers versus high engagers, we see a more similar frequency distribution of average information processing indices in these 2 bins, as compared to contrasting differences in the histogram for in-video dropouts and non dropouts. In order to generalize these findings, we also look at the variations of average IPI among some other student partitions that we made for the whole course. ``Viewers" are students who have watched or interacted with some video lecture but have not done the exercises; the ``Active" students additionally turn in homework also. MOOC dropouts are those students who cease to actively participate in the MOOC (we are concerned with video lecture viewing only) before the last week, i.e, students who do not finish the course. An important observation in figure 5 is that IPI is clearly able to distinguish between Non-dropouts and Dropouts ($|$z$|$=9.06, p$<$0.01). This is also reflected in the histogram in the left half of figure 5, which verifies that more ``Non dropouts" have positive IPI. More is the information processing done by students, greater is the video lecture involvement, higher are the chances to derive true utility from video lecture and remain excited and motivated to stay in the course. We also obtain striking differences between ``Active" versus ``Viewers" ($|$z$|$=10.45, p$<$0.01). Intuitively too, we expect ``Viewers" to have higher IPI than ``Active" class, because as their primary MOOC activity, ``Viewers" grapple more with the video lecture. 
\vspace{-0.55cm}
\section{Conclusion}
\vspace{-0.25cm}
In this work, we have begun to lay a foundation for research investigating students' information processing behavior while interacting with MOOC video lectures. Focusing the center of gravity on the human mind, we applied a cognitive video watching model to explain the dynamic process of cognition involved in MOOC video clickstream interaction. This paved way for the development of a simple, yet potent IPI using linear weight assignments, which can be effectively used as an operationalization for making predictions regarding critical learner behavior. We could contemplate that IPI significantly varies among different student partitions. This actually happens because of presence of smaller substructures inside these larger groupings, that are similar in their click behaviors. Deciphering unique ways of video lecture interaction in such smaller clusters using approaches such as Markov based clustering, would be very meaningful for course instructors, to design customized learning solutions for students within them \cite{SinhaC:14}. It would make sense to incorporate student demographics to better understand some latent factors, such as playback speed choices due to native language differences versus engagement etc. In our recent work \cite{SinhaD:14}, we have been seeking to gain better visibility into how combined representations of video clickstream behavior and discussion forum footprint can provide insights on interaction pathways that lead students to central activities.

\end{document}